\documentclass[
reprint,
superscriptaddress,
 amsmath,amssymb,
 aps,
]{revtex4-2}

\usepackage{graphicx,setspace}
\usepackage{dcolumn}
\usepackage{bm}
\usepackage[mathlines]{lineno}
\usepackage[colorlinks=true,citecolor=teal,linkcolor=olive,anchorcolor=green,urlcolor=teal]{hyperref}
\usepackage{mathrsfs}
\usepackage{xcolor}
\usepackage{float}
\usepackage[normalem]{ulem}
\usepackage[utf8]{inputenc}
\usepackage{tensor}
\usepackage{physics}
\usepackage{blindtext}
\usepackage{slashed}
\usepackage{tikz}
\usetikzlibrary{decorations.pathmorphing}

\newcommand{\be}{\begin{equation}}
\newcommand{\ee}{\end{equation}}
\newcommand{\ba}{\begin{eqnarray}}
\newcommand{\ea}{\end{eqnarray}}

\begin{document}

\title{Higher dimensional Reissner-Nordstr\"{o}m black holes supporting static scalar shells}
\author{Libo Xie}
\email{liboxie@jxnu.edu.cn}
\affiliation{Department of Physics, Jiangxi Normal University, Nanchang 330022, China}
\author{Chaoxi Fang}
\email{chaoxi.f@jxnu.edu.cn}
\affiliation{Department of Physics, Jiangxi Normal University, Nanchang 330022, China}
\author{Jie Jiang}
\email{jiejiang@mail.bnu.edu.cn}
\affiliation{College of Education for the Future, Beijing Normal University, Zhuhai 519087, China}
\author{Ming Zhang}
\email{mingzhang@jxnu.edu.cn (corresponding author)}
\affiliation{Department of Physics, Jiangxi Normal University, Nanchang 330022, China}

\date{\today}

\begin{abstract}
We analytically study scalarization of higher-dimensional charged Reissner-Nordstr\"{o}m (RN) black hole. It is shown that static  massive scalar field which is non-minimally coupled to Gauss-Bonnet invariant can be supported by higher-dimensional black hole in super-critical charge regime $Q/M\ge \bar{C}_d$ with $Q, M$ charge and mass of the  black hole and $\bar{C}_d$ some unitless spacetime dimension-dependent quantity.  Moreover, we show that the static massive scalar shell can be quite thin in the large mass regime $\mu M^{\frac{1}{d-3}}\gg 1$ with $\mu$ mass of the scalar field. 
\end{abstract}
\maketitle

\section{Introduction}
The classical version of no-hair conjecture asserts that the stationary, asymptotically flat black holes in general relativity (GR) and electro-vacuum can be fully characterized  by limited number of macroscopic degrees of freedom--  their conserved mass, angular momentum, and electromagnetic charge \cite{Chrusciel:2012jk,Cardoso:2016ryw}. This conjecture meets  circumvention  beyond GR or even in GR with matter sources beyond electro-vacuum, as additional hairs can be carried, resulting in a much richer landscape of hairy black holes \cite{Herdeiro:2015waa}.

Throughout recent years, the mechanism of spontaneous scalarization, i.e., the destabilization of black holes that are free of scalar and the occurrence of  scalar hair as extra scalar configuration, has got great interest in the scientific community. 
Matter-induced spontaneous scalarization was first raised in neutron star solution of the scalar-tensor theory in the 1990s \cite{Damour:1993hw,AltahaMotahar:2017ijw}, where scalar field is non-minimally coupled to the Ricci curvature. In recent years, curvature-induced spontaneous scalarization (see e.g. \cite{Doneva:2017bvd,Hod:2019pmb,Silva:2017uqg,Antoniou:2017acq,Cunha:2019dwb,Hod:2019vut,Hod:2020jjy}) and charge-induced spontaneous  scalarization (see e.g. \cite{Herdeiro:2018wub,Hod:2019ulh,Hod:2020ius,Hod:2020cal}) due to the tachyonic instability  were discovered and have arose great concerns.
 For the electrovacuum GR  minimally coupled to Gauss-Bonnet (GB) quadratic curvature invariant through a scalar coupling function $f(\phi)>0$, which is dubbed Einstein-Maxwell-scalar-GB (EMsGB) model, spin-induced scalarization were also discovered  \cite{Dima:2020yac,Herdeiro:2020wei,Berti:2020kgk}. Besides, there are other triggers of spontaneous  scalarization \cite{Antoniou:2022agj}, in the context of  a bare mass \cite{Ramazanoglu:2016kul}, different fields \cite{Ramazanoglu:2017xbl}, different coupling functions \cite{Blazquez-Salcedo:2020nhs}, magnetic charges \cite{Annulli:2022ivr}, horizonless reflecting stars \cite{Peng:2019cmm}, non-linear electromagnetic field \cite{Wang:2020ohb}, quasi-topological electromagnetism  \cite{Liu:2019rib,Myung:2020ctt}, asymptotically (A)dS background \cite{Brihaye:2019gla,Guo:2021zed}, etc.
 
 Generically, the evolution equation of the EMsGB theory can only be solved numerically due to the non-linear curvature terms. It has been numerically shown in Refs.  \cite{Brihaye:2019kvj,Herdeiro:2021vjo} that spatially regular nontrivial scalar field  can be supported by the scalarized Reissner-Nordstr\"{o}m (RN) black hole which is differentiated from the bald RN black hole by critical boundary configurations. However, in Ref. \cite{Hod:2022sgu}, infinitesimally thin static scalar shell surrounding the four-dimensional RN black hole in the framework of EMsGB was studied and an analytical resonance spectrum formula as a sharp boundary between bald and hairy RN black hole-massive scalar field configurations  was obtained in large mass (or large coupling) regime. In the present paper, we will mainly follow Ref. \cite{Hod:2022sgu} and extend the study to higher-dimensional case. We will analytically derive the resonance spectrum of the coupling parameter as critical existence-line  characterizing cloudy higher dimensional RN black hole. Our work here can be analytical complement of numerical study of higher dimensional black hole scalarization in Ref. \cite{Astefanesei:2020qxk}.

The outline of our paper is as follows: in Sec. \ref{setup}, we shall derive the dynamic equation describing the massive scalar field non-minimally coupled with the GB invariant; in Sec. \ref{qrs}, we will analytically calculate the quasinormal resonance spectrum of the higher-dimensional black hole-massive scalar field system and show the sharp boundary between the bald higher-dimensional RN black hole and the scalarized higher-dimensional RN black hole solution. Simultaneously, we will give the width of the scalar cloud supported by the black hole; Sec. \ref{conc} will be devoted to our closing remarks.

\section{Setup}\label{setup}

The action of EMsGB field theory we consider is described by the following action,
\begin{equation}\label{act}
\begin{aligned}
S=&\int {\rm{d}}^{d} x \sqrt{-g}\left[\frac{1}{4} R-\frac{1}{4} F_{\alpha \beta} F^{\alpha \beta}-\frac{1}{2} \nabla_{\alpha} \phi \nabla^{\alpha} \phi \right.\\&\left.-\frac{1}{2} \mu^{2} \phi^{2}+f(\phi) \mathcal{G}_d\right],
\end{aligned}
\end{equation}
where $R$ is the Ricci scalar of the spacetime metric $g_{\alpha\beta}$ with determinant $g$, $F_{\alpha\beta}$ is two-form electromagnetic tensor, $\mu$ is  bare mass of the scalar field $\phi$, the GB invariant is
\begin{equation}
\mathcal{G}_d \equiv R_{\mu \nu \rho \sigma} R^{\mu \nu \rho \sigma}-4 R_{\mu \nu} R^{\mu \nu}+R^{2},
\end{equation}
with $R_{\mu\nu}$ the Ricci tensor and $R_{\mu \nu \rho \sigma}$ the Riemann tensor. The coupling function
$f(\phi)$  satisfies 
\begin{equation}
\left.\frac{d f}{d \phi}\right|_{\phi=0}=0.
\end{equation}
 Specifically, the coupling function should share an universal weak-field quadratic behavior,
\begin{equation}\label{cpf}
f(\phi)=\frac{1}{2} \eta \phi^{2}
\end{equation}
for the $d$-dimensional EMsGB field theory \cite{Astefanesei:2020qxk}, with $\eta$ the coupling constant \cite{Astefanesei:2020qxk}. Dimensional analysis implies
\begin{equation}
[R]={\rm{L}}^{-2},\,[\phi]={\rm{L}}^0,\,[\mu]={\rm{L}}^{-1},\, [\eta]={\rm{L}}^{6-d},\,[\mathcal{G}]={\rm{L}}^{-4},
\end{equation}
where $\mathrm{L}$ stands for the dimension of length.

In the spherically symmetric configuration, the coupling function Eq. (\ref{cpf}) guarantees that the bald electrovacuum $d$-dimensional RN spacetime
\begin{equation}\label{met}
d s^{2}=-h(r) d t^{2}+\frac{1}{h(r)} d r^{2}+r^{2} d \theta^{2}+r^{2} d \Omega_{d-2}^{2}
\end{equation}
with the blackening factor being
\begin{equation}
h(r)=1-\frac{16 \pi M r^{3-d}}{(d-2) \epsilon}+\frac{32 \pi^{2} Q^{2} r^{-2(d-3)}}{\left(d^{2}-5 d+6\right) \epsilon^{2}}
\end{equation}
is a scalar-free solution of the theory (\ref{act}) in the weak-field limit. $\epsilon=2 \pi^{\frac{\mathrm{d}-1}{2}} / \Gamma[(d-1) / 2]$ is volume of a $(d-2)$-sphere. The event horizon of the $d$-dimensional RN  black hole (\ref{met}) locates at
\begin{equation}
r_{+}=\left[\frac{4 \pi}{\epsilon}\left(\frac{2 M}{d-2}+\sqrt{\frac{4 M^{2}}{(d-2)^{2}}-\frac{2 Q^{2}}{d^{2}-5 d+6}}\right)\right]^{\frac{1}{d-3}}.
\end{equation}
The dimensions of the black hole's mass $M$ and electric charge $Q$ are
\begin{equation}
[M]={\rm{L}}^{d-3},\,[Q]={\rm{L}}^{d-3}.
\end{equation}

 The Klein-Gordon equation for the scalar field,
\begin{equation}\label{kgeq}
\nabla^{\nu} \nabla_{\nu} \phi=\mu_{\text {eff }}^{2} \phi,
\end{equation}
can be obtained from  the EMsGB action (\ref{act}). We have defined the effective scalar mass as
\begin{equation}
\mu_{\mathrm{eff}}^{2}\equiv\mu^{2}-\eta \mathcal{G}_d,
\end{equation}
where the GB invariant in the $d$-dimensional RN spacetime background (\ref{met}) is specified to be
\begin{equation}
\begin{aligned}
\mathcal{G}_d (r)=&-\frac{4 (d-2) f'(r) f''(r)}{r}-f''(r)^2\\&-\frac{2 (d-3) (d-2) (2 d-7) [f(r)-1]^2}{r^4}\\&-\frac{2 (d-2) (d-1) r f'(r)^2}{r^3}\\&-\frac{8 (d-2)(d-3) f'(r)   [f(r)-1]}{r^3}.
\end{aligned}
\end{equation}
Explicitly, for instance, we have
\begin{equation}
\begin{aligned}
\mathcal{G}_d (r)=
\begin{cases}
&\frac{64\left(72 \pi^{2} M^{2} r^{4}-180 \pi M Q^{2} r^{2}+83 Q^{4}\right)}{9 \pi^{4} r^{12}},\quad d=5,\\
&\frac{27 \left(320 \pi ^2 M^2 r^6-448 \pi  M Q^2 r^3+107 Q^4\right)}{16 \pi ^4 r^{16}},\quad d=6,\\
&\frac{512 \left(300 \pi ^4 M^2 r^8-900 \pi ^2 M Q^2 r^4+437 Q^4\right)}{25 \pi ^8 r^{20}},\quad d=7.
\end{cases}
\end{aligned}
\end{equation}

In the vicinity of the event horizon of the $d$-dimensional RN black hole, the tachyonic condition $\mu_{\mathrm{eff}}^{2}<0$ can be fulfilled in some condition, which signifies the existence of scalar cloud around the black hole. To show that, we perform a decomposition
\begin{equation}\label{deco}
\phi(r, \theta, \Theta)= \sum_{\ell m} \frac{R_{\ell m}(r)}{r^{\frac{d-2}{2}}} Y_{\ell m}(\theta)
\end{equation}
for the static massive scalar field with zero-frequency, where $R_{\ell m}(r)$  and $Y_{\ell m}(\theta, \varphi)$ are respectively the radial function and angular spherical harmonic function with $\ell$, $m$ the spherical and azimuthal harmonic indexes.  Substituting the field decomposition Eq. (\ref{deco})  into the Klein–Gordon equation (\ref{kgeq}) yields a radial  dynamics equation for the massive scalar field,
\begin{equation}\label{rdeq2}
f(r)^{2} R^{\prime \prime}(r)+f^{\prime}(r) f(r) R^{\prime}(r)+U(r) R(r)=0,
\end{equation}
where primes denote $\mathrm{d} / \mathrm{d} r$, and
\begin{equation}
\begin{aligned}
U(r)=&-\frac{(d-4)(d-2) f(r)^{2}}{4 r^{2}} \\
&-\frac{f(r)\left[(d-2) r f^{\prime}(r)+2\left(K_{\ell}+(\mu^{2}-\eta \mathcal{G}_d) r^{2}\right)\right]}{2 r^{2}}
\end{aligned}
\end{equation}
with $K_{\ell}=\ell (d+\ell-3)$ the eigenvalue.

For convenience, after defining a tortoise coordinate $r^*$ by the relation
\begin{equation}\label{tcr}
d r_*=\frac{d r}{h(r)},
\end{equation}
we can transform the radial equation (\ref{rdeq2}) into a Schr\"{o}dinger-like form as
\begin{equation}\label{slf}
\frac{\mathrm{d}^{2} R}{\mathrm{~d} r_*^{2}}-V(r) R=0,
\end{equation}
where the radial effective potential reads
\begin{equation}\label{rdeq}
\begin{aligned}
V(r)=f(r) &\left[\mu^{2}-\eta \mathcal{G}+\frac{K_{\ell}}{r^{2}}+\frac{d^{2}-6 d+8}{4 r^{2}}\right.\\&\left.\quad+\frac{4 \pi(d-2) M}{\epsilon r^{d-1}}-\frac{8 \pi^{2}(3 d-8) Q^{2}}{(d-3) \epsilon^{2} r^{2 d-4}} \right].
\end{aligned}
\end{equation}

To determine the discrete resonance spectrum of the system, physically appropriate static bound-state conditions should be imposed on the  wave function, which are spatially regular massive scalar field  eigenfunction at the event horizon of the black hole and an asymptotically exponential decaying behavior at spatial infinity,
\begin{equation}\label{bc1}
R\left(r=r_{\mathrm{H}}\right)<\infty, \quad R(r \rightarrow \infty) \rightarrow 0.
\end{equation}

The radial differential equation (\ref{slf}) together with the boundary condition Eq. (\ref{bc1}) singles out dimensionless quasinormal resonant frequencies $\left\{M^{\frac{d-6}{d-3}} \eta(M, Q, \mu ; n)\right\}_{n=0}^{n=\infty}$ characterizing the dynamics of the static bound-state massive scalar field nonminimally coupled to the $d$-dimensional RN black hole. Specifically, the fundamental mode ($n=0$) determines a critical existence-line as the boundary between the scalar-free bald $d$-dimensional RN black hole and the scalarized hairy RN black hole.

\section{ quasinormal resonance spectrum of the black hole-massive scalar field system}\label{qrs}
We will show that the $d$-dimensional RN black hole coupled with massive scalar field through the function (\ref{cpf}) where $\eta>0$  is  amenable to analytically determine the quasinormal resonance spectrum in the eikonal large mass regime
\begin{equation}\label{elmr}
\mu\left[\frac{8 \pi}{(d-2) \epsilon} M\right]^{\frac{1}{d-3}} \gg d+\ell-3,
\end{equation}
which is equivalent to a dimensionless large-coupling regime
\begin{equation}\label{lcr}
\bar{\eta} \equiv \frac{\eta}{M^{\frac{6-d}{d-3}}} \gg 1.
\end{equation}

Accordingly, we will analytically calculate the quasinormal resonance spectrum of the $d$-dimensional RN black hole-massive scalar field system in the large-coupling regime (\ref{lcr}), where the radial effective potential (\ref{rdeq}) can be reduced to
\begin{equation}\label{effr}
\begin{aligned}
V(r , \bar{\eta})=&f(r)\left[\mu^{2}-\bar{\eta} V_{\mathrm{GB}}(r)\right] \\&\times\left\{1+\mathcal{O}\left[\left(\mu M^{\frac{1}{d-3}}\right)^{-2}\right]\right\},
\end{aligned}
\end{equation}
where the redefined dimensional GB invariant term is
\begin{equation}
V_{\mathrm{GB}}(r )=M^{\frac{6-d}{d-3}}\mathcal{G}_d (r).
\end{equation}

Through the relation $\mathcal{G}_d^{\prime}(r)=0$, we can know that the redefined GB invariant peaks at 
\begin{equation}\label{rgip}
r_{\text {p}}(M, Q)=C_d\left(\frac{Q^{2}}{M}\right)^{\frac{1}{d-3}},
\end{equation}
where $C_d$ is some dimension-dependent dimensionless quantity; for example, for $d=5$, we have $C_5=\frac{1}{2} \sqrt{\frac{1}{6\epsilon} \times(75+\sqrt{1641}) \pi}$. The critical condition that the peak coincides with the event horizon of the black hole is 
\begin{equation}
\frac{Q}{M} =\left(\frac{Q}{M}\right)_{\text {crit }}=\bar{C}_d,
\end{equation}
where $\bar{C}_d$ is one other $d$-dependent dimensionless quantity. For example, we have
\begin{equation}\label{barcd}
\bar{C}_d=\begin{cases}\frac{8}{83} \sqrt{108-\sqrt{\frac{547}{3}}}=0.93696,\quad d=5, \\ \frac{1}{107} \sqrt{12393-48 \sqrt{2931}}=0.92492,\quad d=6, \\  \frac{4}{437} \sqrt{13410-462 \sqrt{\frac{283}{5}}}=0.91232,\quad d=7.\end{cases}
\end{equation}
Thus we get the maximal value of the redefined GB invariant  as
\begin{equation}
\begin{aligned}
\max _{r} &\left\{V_{\mathrm{GB}}\left(r \geq r_{+} \right)\right\} \\
&= \begin{cases}\frac{M^{\frac{3d-2}{d-3}}}{Q^{\frac{4(d-1)}{d-3}}} X_d,  & \text { for } Q / M \geq \bar{C}_d, \\ V_{\mathrm{GB}}(r=r_+ ), & \text { for } Q / M \leq \bar{C}_d,\end{cases}
\end{aligned}
\end{equation}
where
\begin{equation}
\begin{aligned}
&X_d=\frac{2048 \pi ^4 \left[d \left(4 d^2-26 d+53\right)-32\right] C_d^{8-4 d}}{(d-3) (d-2)^2 \epsilon ^4}\\&+\frac{256 \pi ^2 C_d^{2-3 d} \left[(d-3) (d-1) \epsilon  C_d^d+8 \pi  (5-2 d) C_d^3\right]}{\epsilon ^3}
\end{aligned}
\end{equation}
is some dimensionless quantity dependent on the spacetime dimension $d$.

The necessary condition for the existence of the massive bound-state scalar cloud well outside the event horizon of the $d$-dimensional RN black hole is that the effective potential should be attractive, that is, we require
\begin{equation}
V\left(r_{t_{-}} \leq r \leq r_{t_{+}}\right) \leq 0 \quad \text { with } \quad r_{t_{-}} \geq r_{+},
\end{equation}
where $\left\{r_{t_{-}}, r_{t_{+}}\right\}$ are two turning points where $V(r_{t_{\pm}} , \bar{\eta})=0$. Thus we have a series of inequalities,
\begin{equation}
\begin{aligned}
\mu^{2}-\bar{\eta} \cdot \max _{r}\left\{V_{\mathrm{GB}}(r)\right\} & \leq \mu^{2}-\bar{\eta} \cdot V_{\mathrm{GB}}(r) \\
& \leq \mu^{2}+\frac{K_{l}}{r^{2}}+\frac{d^{2}-6 d+8}{4 r^{2}}\\ &\quad+\frac{4 \pi(d-2) M}{\epsilon r^{d-1}}\\&\quad-\frac{8 \pi^{2}(3 d-8) Q^{2}}{(d-3) \epsilon^{2} r^{2 d-4}}-\bar{\eta} \cdot V_{\mathrm{GB}}(r) \\&\leq 0,
\end{aligned}
\end{equation}
which gives the upper bound of the bald mass of the scalar cloud as
\begin{equation}\label{ubb}
\begin{aligned}
M^{\frac{1}{d-3}} \mu \leq \sqrt{\bar{\eta}} \cdot \begin{cases}\sqrt{X_d}\cdot \frac{M^{\frac{3d}{2(d-3)}}}{Q^{\frac{2(d-1)}{d-3}}}, \,\, \text { for } Q / M \geq \bar{C}_d, \\  M^{\frac{1}{2(d-3)}}V_{\mathrm{GB}}(r_+ ), \,\, \text { for }Q / M \leq \bar{C}_d.\end{cases}
\end{aligned}
\end{equation}
This relation, on the other hand, also implies the lower bound of the dimensionless coupling parameter,
\begin{equation}\label{ubbm}
\begin{aligned}
\bar{\eta}  \geq M^{\frac{2}{d-3}} \mu^2 \cdot \begin{cases}X_d^{-1}\cdot \frac{Q^{\frac{4(d-1)}{d-3}}}{M^{\frac{3d}{d-3}}},\,\, m  \text { for } Q / M \geq \bar{C}_d, \\  M^{\frac{1}{3-d}}V_{\mathrm{GB}}^{-2}(r_+ ),\,\, \text { for }Q / M \leq \bar{C}_d.\end{cases}
\end{aligned}
\end{equation}

Now we further focus on the  super-critical charge regime
\begin{equation}
\frac{Q}{M} \geq \bar{C}_d,
\end{equation}
where we will explicitly  show that the higher dimensional RN black hole is endowed with an arbitrarily thin scalar shell hovering a finite proper distance. We shall analytically obtain the quasinormal resonance spectrum  $\left\{M^{\frac{d-6}{d-3}} \eta(M, Q, \mu ; n)\right\}_{n=0}^{n=\infty}$ characterizing the system composed of nonminimally coupled massive scalar field and higher-dimensional RN black hole in the eikonal large mass regime Eq. (\ref{elmr}) which is equivalent to the large-coupling regime Eq. (\ref{lcr}). To that end, we use the second-order WKB  quantization condition  \cite{bender1999advanced} 
\begin{equation}
\int_{y_{t_{-}}}^{y_{t_{+}}} d y \sqrt{-V(y, \bar{\eta})}=\left(n+\frac{1}{2}\right) \cdot \pi, \quad n=0, 1, 2 \cdots
\end{equation}
which characterizing Schr\"{o}dinger-like equation Eq. (\ref{slf}) with a reduced effective potential Eq. (\ref{effr}). The two integration limits $y_{t_{-}}$, $y_{t_{+}}$ are turning points of the effective potential which make $V\left(y_{t_{-}}\right)=V\left(y_{t_{+}}\right)=0$. This equation can be transformed into a more mathematically amenable form
\begin{equation}\label{sleq}
\int_{r_{t_{-}}}^{r_{t_{+}}} d r \sqrt{-\frac{V(r, \bar{\eta})}{[h(r)]^{2}}}=\left(n+\frac{1}{2}\right) \cdot \pi, \quad n=0,1,2 \cdots
\end{equation}
by applying the definition of the tortoise coordinate Eq. (\ref{tcr}).

Following the suggestion of Ref. \cite{Hod:2022sgu}, we here define two auxiliary dimensionless variables $\chi$ and $x$ by relations [cf. Eqs. (\ref{rgip}) and (\ref{ubb})]
\begin{equation}\label{adv1}
M^{\frac{1}{d-3}} \mu =\sqrt{\bar{\eta}} \sqrt{X_d}\cdot \frac{M^{\frac{3d}{2(d-3)}}}{Q^{\frac{2(d-1)}{d-3}}} \cdot(1-\chi), \quad \quad \chi \geq 0,
\end{equation}
\begin{equation}
r = r_{\text {p }} (1+x).
\end{equation}
Then the radial effective potential (\ref{effr}) can be reduced to
\begin{equation}\label{rrep}
\begin{aligned}
M^{\frac{2}{d-3}} \frac{V(r)}{h(r)^{2}}&=\frac{M^{\frac{2}{d-3}}\left[\mu^{2}-\bar{\eta} V_{\mathrm{GB}}(r )\right] }{h(r)}\\&=-\frac{2 M^{\frac{2}{d-3}} \tilde{\mu} _m^2}{\frac{32 \pi ^2 Q^2 r_{\text {p }}^{6-2 d}}{\left(d^2-5 d+6\right) \epsilon ^2}-\frac{16 \pi  M r_{\text {p }}^{3-d}}{(d-2) \epsilon }+1}\\&\quad\times\left(\mathfrak{X}_d x^2+\chi\right)\left[1+\mathcal{O}(x, \chi)\right],
\end{aligned}
\end{equation}
where $\tilde{\mu}_{m}\equiv \sqrt{\bar{\eta}X_d}M^{\frac{2-3 d}{6-2 d}} Q^{-\frac{2 (d-1)}{d-3}}$, and $\mathfrak{X}_d<0$ is a dimension-dependent dimensionless constant. For example, 
\begin{equation}
\mathfrak{X}_d=\begin{cases}\frac{3}{59} \left(5 \sqrt{1641}-547\right),\quad d=5, \\ -28.8207,\quad d=6, \\  \frac{30}{119} \left(3 \sqrt{1415}-283\right),\quad d=7.\end{cases}
\end{equation}

Combining Eqs. (\ref{sleq}) and (\ref{rrep}), we find that the discrete quasinormal resonance spectrum of the black hole-massive scalar field system is

\begin{equation}\label{dqrs}
\begin{aligned}
&\chi (\bar{\eta})=(2 n+1)\\&\times \sqrt{\frac{\mathfrak{X}_d Q^{\frac{4 (d-1)}{d-3}} \left(\frac{8 \pi  r_\mathrm{p}^{3-2 d} \left((d-3) M \epsilon  r_\mathrm{p}^d-2 \pi  Q^2 r_\mathrm{p}^3\right)}{(d-3) (d-2) \epsilon ^2}+\frac{1}{2}\right)}{\bar{\eta } M^{\frac{3 d-2}{d-3}} X_d r_\mathrm{p}^2}},\\&\quad n=0,1,2 \cdots
\end{aligned}
\end{equation}
In the large-coupling regime $\bar{\eta}\gg 1$, we have
\begin{equation}
\chi (\bar{\eta}) \ll 1
\end{equation}
for the characteristic parameter. Furthermore, substituting this resonance spectrum Eq. (\ref{dqrs}) into Eq. (\ref{adv1}), we get an explicit resonance formula
\begin{equation}\label{erft}
\begin{aligned}
\sqrt{\bar{\eta}}=&\frac{\mu  M^{\frac{2-3 d}{2 (d-3)}} Q^{\frac{2 (d-1)}{d-3}}}{\sqrt{X_d}}+(2 n+1)\\&\times \sqrt{\frac{\mathfrak{X}_d  Q^{\frac{4 (d-1)}{d-3}} \left(\frac{8\pi  r_\mathrm{p}^{3-2 d} \left(2 \pi  Q^2 r_\mathrm{p}^3-(d-3) M \epsilon  r_\mathrm{p}^d\right)}{(d-3) (d-2) \epsilon ^2}+\frac{1}{2}\right)}{M^{\frac{3 d-2}{d-3}}X_d r_\mathrm{p}^2}},\\&\quad n=0,1,2 \cdots
\end{aligned}
\end{equation}

We shall continue to calculate the effective width of the massive scalar field supported by the $d$-dimensional RN black hole, which can be roughly defined by the turning points of the radial effective potential,
\begin{equation}\label{trep}
\Delta r(Q / M, \mu M^{\frac{1}{3-d}}) \equiv r_{t_{+}}-r_{t_{-}}.
\end{equation}
To this end, we define 
\begin{equation}\label{bdef}
\mathcal{X}=\sqrt{-\frac{\mathfrak{X}_d}{\chi}}x,
\end{equation}
then the WKB  quantization condition (\ref{sleq})  can be transformed to
\begin{equation}\label{wkbqc}
\int_{-1}^{1}\chi r_\mathrm{p}\sqrt{\frac{\mathcal{P}}{\mathfrak{X}_d}(1-\mathcal{X}^2)}dz=\left(n+\frac{1}{2}\right)\pi,
\end{equation}
where
\begin{equation}
\mathcal{P}\equiv -\frac{2\bar{\eta}X_d M^{\frac{2-3d}{3-d}}Q^{\frac{4(d-1)}{3-d}}}{{\frac{32 \pi ^2 Q^2 r_{\text {p }}^{6-2 d}}{\left(d^2-5 d+6\right) \epsilon ^2}-\frac{16 \pi  M r_{\text {p }}^{3-d}}{(d-2) \epsilon }+1}}.
\end{equation}

Using Eqs. (\ref{rgip}), (\ref{bdef}) and (\ref{wkbqc}), we get a reduced dimensionless effective width of the cloud composed of massive scalar field supported by the central $d$-dimensional RN black hole as
\begin{equation}\label{rdew}
\Delta r M^{\frac{1}{3-d}}=2C_d \left(\frac{Q}{M}\right)^{\frac{2}{d-3}}\sqrt{\frac{\chi}{\mathfrak{X}_d}},
\end{equation}
which in the large-coupling regime can be further transformed to
\begin{equation}\label{woc}
\begin{aligned}
&\Delta r M^{\frac{1}{3-d}}=\sqrt{\frac{2 \sqrt{2} (2 n+1) C_d M^{\frac{4}{3-d}} Q^{\frac{6}{d-3}} }{\mu  M^{\frac{1}{d-3}}}}\\&\times \left(\frac{16 \pi  r_\mathrm{p}^{3-2 d} \left[2 \pi  Q^2 r_\mathrm{p}^3-(d-3) M \epsilon  r_\mathrm{p}^d\right]}{(d-3) (d-2) \mathfrak{X}_d^2 \epsilon ^2}+\frac{1}{\mathfrak{X}_d}\right)^{1/4}
\end{aligned}
\end{equation}
by substituting Eqs. (\ref{dqrs}), (\ref{erft}) into Eq. (\ref{rdew}). Strikingly, in the large mass regime Eq. (\ref{elmr}), the width of the scalar field can be quite thin.

Moreover, in the eikonal large mass regime we have
\begin{equation}
\sqrt{V} \propto M^{\frac{1}{d-3}} \mu
\end{equation}
for the radial effective potential according to Eqs.  (\ref{effr}) and (\ref{erft}). Consequently,  outside the region given by Eq. (\ref{trep}),  i.e., in the classically inaccessible region, the massive scalar field is exponentially damped with an asymptotic form
\begin{equation}
R=\text { constant } \times e^{-\int \sqrt{V(y)} d y}\sim e^{M^{\frac{1}{d-3}} \mu \mathfrak{d}}.
\end{equation}
As $M^{\frac{1}{d-3}} \mu\gg 1$, we just have the penetration depth
\begin{equation}
\mathfrak{d}\sim \frac{1}{M^{\frac{1}{d-3}} \mu}\ll 1
\end{equation}
for the massive scalar field supported by the central $d$-dimensional RN black hole.

\section{Closing remarks}\label{conc}
In the present paper, we studied the spontaneous scalarization of higher-dimensional charged RN black hole, triggered by massive scalar field non-minimally coupled with GB curvature invariant.  We derived the 
Schr\"{o}dinger-like differential equation (\ref{slf}) describing the radial dynamics of the massive scalar field that is characterized by a radial effective potential (\ref{rdeq}). In the eikonal large mass regime (\ref{elmr}), or equivalently in the large coupling regime (\ref{lcr}), the radial effective potential tends to be Eq. (\ref{effr}). We then obtain a coupling parameter-dependent upper bound (\ref{ubb}) for the mass of the scalar field supported by the scalarized higher dimensional RN black hole, which increases with the  increasing reduced coupling parameter $\bar{\eta}$ both in the super-critical and sub-critical charge regimes.

Furthermore, we focused on the super-critical charge regime, where the peak of the radial effective potential just locates outside the event horizon of the $d$-dimensional RN black hole, we analytically derived the resonance spectrum formula Eq. (\ref{erft}) for the massive scalar field using WKB method. This compact formula just manifests the critical boundary between the scalar-free bald higher-dimensional RN black hole solution and the scalarized higher-dimensional RN black hole solution for the EMsGB field theory (\ref{act}). Quite strikingly,  after explicitly obtained the width formula Eq. (\ref{woc}) for the scalar cloud supported by the hairy higher-dimensional RN black hole,  we found that this scalar shell in the large coupling regime can be quite thin. 

We note that, similar to the four-dimensional case shown in Ref. \cite{Hod:2022sgu}, in the subcritical charge regime with a positive coupling parameter $\eta$, thin static massive scalar field can also be supported by the higher-dimensional RN black hole. However, in that case, the scalar configuration is connected to the event horizon of the black hole as $r_\mathrm{p}<r_+$. Additionally, it is not difficult to understand that even in the regime with a negative coupling parameter, the EMsGB theory contain scalarized higher-dimensional RN black hole that supports static thin massive scalar shell \cite{Herdeiro:2021vjo}.

Early exploration in scalarization of charged black holes in the Einstein-Maxwell-scalar model can be seen in Refs. \cite{Gubser:2005ih,Gubser:2008px}. As elaborated in Ref. \cite{Astefanesei:2019pfq}, there exist two kinds of black holes in the theory: the dilatonic and the scalarized ones. It was shown that  a purely electrically ( or magnetically) charged  black  hole in the theory does not have a regular extremal limit, but instead a critical one; in contrast, dyonic black holes can have regular extremal limits both for the dilatonic and scalarized black holes \cite{Astefanesei:2019pfq}. In this paper, the RN black hole in the extremal limit has vanishing temperature but non-vanishing entropy and the spacetime is regular both on and outside its smooth event horizon. It also has a symmetry enhanced near-horizon geometry of $A d S_{2} \times S^{2}$ \cite{robinson1959solution,Bertotti:1959pf}. The near-extremal limit of $d$-dimensional RN black hole can be attained when $Q^2\to 2(d-3)M^2/(d-2)$. In this limit, according to Eqs. (\ref{erft}) and (\ref{woc}), it is of interest that the existence curve between the bald RN black hole and the scalarized one  still survives and we can have a near-extremal RN black hole supporting a thin massive scalar field shell (note that weather the extremal limit can be attained relies on specific numerical solution of the scalarized black hole \cite{Doneva:2018rou}). This is consistent with the well-known fact that the singular horizon can be dressed by the coupling of the scalar field to higher derivative terms.  In this sense, the result partially extends the one for massless scalar field supported by four-dimensional RN black hole reported in Ref. \cite{Brihaye:2019kvj}.

Moreover, comparing with the four-dimensional case in Ref. \cite{Hod:2022sgu}, here in the higher dimensional cases, it is worth noting a point that the super-critical charge regime parameter $\bar{C}_d$ which makes the radial peak of the GB invariant (\ref{rgip}) coincide with the event horizon has a nontrivial change from $d=4$ to $d\geq 5$.  That is, we have $\bar{C}_4=0.91652$ \cite{Hod:2022sgu}; it increases to $\bar{C}_5=0.93696$ but then decreases with the increasing spacetime dimension $d$ (this is not relevant to whether $d$ is even or odd, as we have $\bar{C}_8=0.90198$, cf. (\ref{barcd})).  Thermodynamically, RN black holes have stable near-extremal phases in  canonical ensemble with fixed electric charge \cite{Chamblin:1999tk}, but are unstable in  grand canonical ensemble with fixed electric potential. Dynamically, we in this paper showed that it can hold static bound-state scalar field configurations in general spacetime dimensions. This can be attributed to the non-trivial coupling of the massive scalar field with the higher derivative GB term, which provides a binding effective potential as a finite box outside the black hole's event horizon. This is quite similar to the case studied in Ref. \cite{Astefanesei:2019qsg} for  Einstein-Maxwell-dilaton black holes \cite{Anabalon:2013qua}, which can be viewed as an interpolated solution between RN and the so-called Gibbons-Maeda-Garfinkle-Horowitz-Strominger solution \cite{Gibbons:1987ps,Garfinkle:1990qj}. There, it was shown that the dilaton potential behaving as $\alpha \phi^{5} / 30$ for small dilaton field $\phi$ with $\alpha$ a dimensionful  constant contributes to the  effective potential, yielding a regular near-horizon geometry in the extremal regime and making it possible for the black hole to be both thermodynamically stable in canonical/grand canonical ensembles and dynamically stable against linear/non-linear  perturbations.

\section*{Acknowledgements}
M. Z. is supported by the National Natural Science Foundation of China with Grant No. 12005080.  J. J. is supported by the Guangdong Basic and Applied Research Foundation with Grant No. 217200003 and the Talents Introduction Foundation of Beijing Normal University with Grant No. 310432102.

\bibliography{refs}
\end{document}